\documentclass{article}

\usepackage[margin=1in]{geometry}
\usepackage[utf8]{inputenc}
\usepackage{amsmath}
\usepackage{amssymb}
\usepackage{amsthm}
\usepackage{xargs}
\usepackage{bm}
\usepackage{graphicx}
\usepackage{booktabs}
\usepackage{tabularx}
\usepackage[colorlinks=true, linkcolor=blue]{hyperref}
\usepackage{scalerel,stackengine}
\usepackage{xcolor}
\usepackage{authblk}
\usepackage{multirow}
\usepackage{dsfont}

\usepackage[numbers]{natbib}
\newtheorem{theorem}{Theorem}[]

\theoremstyle{definition}

\theoremstyle{remark}

\newcommand{\numeral}[1]{%
  \textup{\uppercase\expandafter{\romannumeral#1}}%
}

\stackMath
\newcommand\reallywidehat[1]{%
\savestack{\tmpbox}{\stretchto{%
  \scaleto{%
    \scalerel*[\widthof{\ensuremath{#1}}]{\kern.1pt\mathchar"0362\kern.1pt}%
    {\rule{0ex}{\textheight}}
  }{\textheight}%
}{2.4ex}}%
\stackon[-6.9pt]{#1}{\tmpbox}%
}
\title{
Designing efficient randomized trials: power and sample size calculation when using semiparametric efficient estimators
}

\author[1]{Alejandro Schuler\thanks{aschuler@unlearn.ai}}
\affil[1]{Unlearn.AI, Inc., San Francisco, California}

\author[ ]{\\{\footnotesize for the Critical Path for Alzheimer's Disease, }\thanks{Data used in the preparation of this article were obtained from the Critical Path Institute's Critical Path for Alzheimer's Disease (CPAD) consortium. As such, the investigators within CPAD contributed to the design and implementation of the CPAD database and/or provided data, but did not participate in the analysis of the data or the writing of this report.}}
\author[ ]{{\footnotesize the Alzheimer's Disease Neuroimaging Initiative, }\thanks{Data used in preparation of this article were obtained from the Alzheimer’s Disease Neuroimaging Initiative (ADNI) database (\url{adni.loni.usc.edu}). As such, the investigators within the ADNI contributed to the design and implementation of ADNI and/or provided data but did not participate in analysis or writing of this report. A complete listing of ADNI investigators can be found at \url{http://adni.loni.usc.edu/wp-content/uploads/how_to_apply/ADNI_Acknowledgement_List.pdf}}}
\author[ ]{{\footnotesize and the Alzheimer's Disease Cooperative Study}\thanks{Data used in preparation of this manuscript/publication/article were obtained from the University of California, San Diego Alzheimer’s Disease Cooperative Study. Consequently, the ADCS Core Directors contributed to the design and implementation of the ADCS and/or provided data but did not participate in analysis or writing of this report.}}

\date{\today}

\begin{document}

\newcommandx{\E}[2][1]{\mathbb E_{#1} \left[#2\right]}
\newcommand{\V}[1]{{\mathbb V} \left[#1\right]}
\newcommand{\C}[1]{{\mathbb C} \left[#1\right]}
\newcommand{\iid}[0]{\overset{\text{IID}}{\sim}}
\newcommandx{\Ehat}[2][1]{\widehat{\mathbb E}_{#1} \left[#2\right]}
\newcommandx{\Vhat}[2][1]{\widehat{\mathbb V}_{#1} \left[#2\right]}
\newcommandx{\rightarrowp}[0]{\overset{p}{\rightarrow}}

\maketitle
\begin{abstract}
Trials enroll a large number of subjects in order to attain power, making them expensive and time-consuming. Sample size calculations are often performed with the assumption of an unadjusted analysis, even if the trial analysis plan specifies a more efficient estimator (e.g. ANCOVA). This leads to conservative estimates of required sample sizes and an opportunity for savings. Here we show that a relatively simple formula can be used to estimate the power of any two-arm, single-timepoint trial analyzed with a semiparametric efficient estimator, regardless of the domain of the outcome or kind of treatment effect (e.g. odds ratio, mean difference). Since an efficient estimator attains the minimum possible asymptotic variance, this allows for the design of trials that are as small as possible while still attaining design power and control of type I error. The required sample size calculation is parsimonious and requires the analyst to provide only a small number of population parameters. We verify in simulation that the large-sample properties of trials designed this way attain their nominal values. Lastly, we demonstrate how to use this formula in the ``design'' (and subsequent reanalysis) of a real randomized trial and show that fewer subjects are required to attain the same design power when a semiparametric efficient estimator is accounted for at the design stage.

\end{abstract}

\section{Introduction}

Clinical research often aims to estimate the effect of a treatment on an outcome of interest \cite{maldonado}. The randomized trial is the gold standard for causal inference because randomization cancels out the effects of any unobserved confounders in expectation \cite{Sox:2012hu, Overhage:2013fx, Hannan:2008gh}. Randomization, however, does nothing to combat the natural variability that comes with finite samples of a larger population. Because of this, treatment effect estimates from a trial are always accompanied by a measure of uncertainty. The expected degree of uncertainty determines how likely the trial is to positively identify an effect of a certain size (the \textit{power} of the trial). Since trials are expensive and time-consuming, it is standard to design for power of $\ge$80\% or 90\% to minimize the likelihood of failure.

Several factors affect power. In trials where subjects are assumed independent these factors separate into characteristics of the data-generating process (population, disease, etc.), the aggressiveness of the rule used to determine ``success'' (e.g. the p-value cutoff), the number of subjects enrolled into the trial (trial size), and the method of data analysis \cite{jones-power}. The population of interest is determined by clinical desiderata and is therefore outside of the statistician's control. The decision rule is tied to a desired type I error rate (false positive rate) and so is typically fixed. This leaves trial size and the analysis method as the primary determinants of power that may be modified.

For example, consider a trial of $n$ subjects with 1:1 randomization. This study is to determine the effect that a new drug has on blood pressure for hypertensive patients after a year of treatment. The treatment effect here is the theoretical difference in blood pressure on treatment vs. placebo averaged across the entire population of hypertensive patients. To estimate the effect using our trial data we will take the difference of the mean blood pressure after one year in the control and treatment groups and test for the effect using an unpaired z-test (two-sided). This is a canonical ``unadjusted'' analysis. Assuming that the variability in the outcome is the same in each treatment arm ($\sigma^2$), it is known that the sampling variance of the effect estimated in this fashion will be near $4\sigma^2/n$ in large-enough samples (where $n$ is the size of the trial) \cite{procova}. Basic statistical theory establishes that the power of this trial to detect an effect of size $\tau$ at significance level $\alpha$ is therefore 
$
\Phi
        \left(
            \Phi^{-1}(\alpha/2) + 
            \sqrt{n}\frac{\tau}{2\sigma}
        \right) 
    +
    \Phi
        \left(
            \Phi^{-1}(\alpha/2) -
            \sqrt{n}\frac{\tau}{2\sigma}
        \right)
$
where $\Phi$ is the standard normal cumulative distribution function. 

In order to achieve a trial with power greater than, say, 80\%, we must increase $n$ until the result of the above formula exceeds 0.8. This determines the necessary enrollment for the trial. Formulas such as the one above (or approximations thereof) are routinely used to determine the size of a trial before it is begun \cite{jones-power, dupont-power}. The form of the calculation depends on the estimator and test used \cite{dupont-power}. We are using a simple unadjusted estimate and the corresponding z-test in our example so the formula turns out to be relatively simple. Note that the only population characteristic we need to assume (or estimate from prior data) is the outcome variability $\sigma$.

Given a target effect, the sampling variance of the estimated effect drives power, regardless of the specific formula associated with a given estimator. The smaller the sampling variance of an estimator, the more power a trial will have. This motivates the use of estimators that have less theoretical sampling variance (are more \textit{efficient}). For a completed trial, using a more efficient estimator would typically result in smaller confidence intervals and smaller p-values. For example, adjusted linear regression (ANCOVA) is typically more efficient\footnote{
And is \textit{guaranteed} to be if treatment-covariate interactions are included in the adjustment or if randomization is 1:1. \cite{leon}
}
than difference-in-means estimation and so is often used in the analysis of trial data even when the trial is powered under the assumption of a difference-in-means analysis \cite{turner}. Unfortunately, however, it is difficult to customize power calculations to adjusted estimators without specifying or assuming a large number of population parameters \cite{sample-logistic, sample-logistic2, sample-logistic3, sample-poisson}.

There is a statistical limit to efficiency in the estimation of treatment effects. That is, without making unverifiable assumptions about the data-generating process, there is a sampling variance that no reasonable estimator can beat. But there are often feasible estimators that attain this limit. These estimators are called \textit{semiparametric efficient}. By definition, no estimator is better than these semiparametric estimators in terms of asymptotic sampling variance \cite{Tsiatis:2007vl}. The benefit of using a semiparametric efficient estimator is that the resulting confidence intervals will be as small as possible while still being valid in general settings (e.g. without assumptions of linearity, homoscedasticity, etc.). This in turn implies maximal power while maintaining control of type I error.

Semiparametric efficient estimators have been previously used in the (re)analysis of trial data and related simulations, where they have been empirically shown to be effective tools for increasing confidence (and thus power) while controlling type I error \cite{knaus-dml, yang-dml-applied, moore-tmle-binary-trials, zhang-dml-trials, zivich, rothe-rct, Wager:2016dz}. But to our knowledge it has not been demonstrated how to leverage their beneficial properties in order to benefit trial \textit{design}. That is our task in this paper. In other words, instead of using more efficient estimators in order to shrink uncertainty for a trial with fixed design, we show how to use them to prospectively design smaller trials that maintain their power. We do this without requiring the estimation or assumption of a large number of population parameters. The practical benefits of this are immense. Each patient in the trial contributes substantial cost and duration, so smaller trials mean cheaper and faster development cycles for treatments that patients need.

Our primary contribution is the development and exposition of a practical power formula that pertains to any semiparametric efficient estimator used in a trial, regardless of the outcome type and (marginal) treatment effect estimand. We verify in simulation that the large-sample properties of trials designed this way attain their nominal values. Lastly, we demonstrate how to use this formula in the ``design'' (and subsequent reanalysis) of a real randomized trial and show that fewer subjects are required to attain the same design power when a semiparametric efficient estimator is accounted for at the design stage.

\section{Background}

\subsection{Setting and Notation}
Our setting is a two-arm randomized controlled trial (RCT) with a single-timepoint outcome. Denote the outcome for subject $i$ in the randomized trial with $Y_i$, their baseline covariates with $X_i$, and their treatment assignment with $W_i$. The trial dataset is a set of $n$ tuples $(X_i, W_i, Y_i)$, which we denote $(\bm X, \bm W, \bm Y) \in \mathcal X^n \times \{0,1\}^n \times \mathbb R^n$ (we use boldface $\bm A$ to denote a vector of random variables, each associated with one observation in the dataset). Let $Y_0$ and $Y_1$ be the control and treatment potential outcomes of the subjects in the trial, respectively, and let $\bm Y_{\bm W} = \bm W \bm Y_1 + (1-\bm W) \bm Y_0$ \cite{Rubin2005}. Our structural assumption about the trial is,
\begin{equation}
 P(\bm X, \bm W, \bm Y, \bm Y_0, \bm Y_1) = 
 \bm 1(\bm Y = \bm Y_{\bm W})
 P(\bm W) 
 \prod_i P(X_i, Y_{0,i}, Y_{1,i}) \, .
 \label{rct-structure}
\end{equation}
In other words, a) the observed outcomes are the potential outcomes corresponding to the assigned treatment, b) the treatment is assigned independently of everything (thus at random), and c) our counterfactual trial subjects are independent of each other. In addition to being independent, we also assume the subjects are identically distributed, i.e., $(X_i, Y_{0,i}, Y_{1,i}) \iid P(X, Y_{0}, Y_{1})$. 

Denote the marginal mean outcomes under each treatment condition $w$ as $\mu_w = \E{Y_w}$ and denote the conditional means $\mu_w(X) = \E{Y_w|X}$. Our structural assumptions about the trial ensure that $\E{Y|W=w} = \E{Y_w}$ and $\E{Y|X, W=w} = \E{Y_w|X}$, so the collected data can be used to make inferences about counterfactual quantities of interest.  In general, a (marginal) ``treatment effect'' is any function of the marginal means, i.e. $\tau = r(\mu_0, \mu_1)$. Examples of this are the difference in means $\tau = \mu_1 - \mu_0$ and odds ratio $\tau = (\mu_1/(1-\mu_1))/(\mu_0/(1-\mu_0))$. For the purposes of this paper we also require that $\frac{\partial r}{\partial \mu_0} \leq 0$ and $\frac{\partial r}{\partial \mu_1} \geq 0$. This sensible condition is satisfied by most definitions of the treatment effect, including difference-in-means and odds ratio, and essentially means that the treatment effect is increasing in $\mu_1$ and decreasing in $\mu_0$ as would logically be expected from a definition of a treatment effect.

Finally, we denote the treatment indicators as $W_{1,i} = W_i$ and $W_{0,i} = 1-W_i$ to allow for symmetric notation. In other words, $W_w = 1(W=w)$ for the random variable $W$ and the constant $w$. Let $\pi_{1} = P(W_{1}=1)$ and let $\pi_0 = P(W_{0}=1)$ be the probability that a subject is assigned to the treatment or control arm in the trial, respectively. In simple randomized experiments, these are constants that apply to all subjects.

In what follows, we abbreviate the usual empirical (sample) average of IID variables $A_1 \dots A_n \sim A$ with the notation $\Ehat{A} = \frac{1}{n} \sum A_i$ (or $\bar A$).
Denote an empirical \textit{conditional} average $\Ehat{A|B=b} = \frac{1}{n_b} \sum_{B_i=b} A_i$ with $n_b$ the number of observations where $B_i=b$. 
Let $\tilde A = A - \E A$ (or $\tilde A = A - \Ehat A$) be centered (or empirically centered) versions of the random variable $A$, with usage clear from context or otherwise noted. Let $\V{A}$ denote the variance of $A$ and $\C{A,B}$ denote the covariance between $A$ and $B$. 
When we describe ``asymptotic'' properties of an estimator in all cases we are referring to the asymptote where the number of observations is increasing while other properties of the data-generating process remain fixed.

\subsection{Semiparametric efficient estimators}
\label{sec:estimator}

No ``reasonable''\footnote{regular and asymptotically linear- see appendix for details.} estimator can possibly attain an asymptotic variance lower than that of a semiparametric efficient estimator \cite{Tsiatis:2007vl}. The benefit of using a semiparametric efficient estimator is therefore that the resulting confidence intervals will be as small as possible while still being valid in general settings (e.g. without assumptions of linearity, homoscedasticity, etc.). This in turn implies maximal power while maintaining control of type I error.

A few semiparametric efficient estimators for marginal treatment effects have been described in the literature \cite{Chernozhukov:2018fb, vanderLaan:2011ep, diaz-tmle-dml}. The differences between these estimators pertain to their small-sample properties or the technical assumptions they require to attain efficiency, neither of which are relevant to our current discussion. To fix ideas we will describe one semiparametric efficient estimator in detail here, but our subsequent discussion applies equally to any estimator that attains the semiparametric efficiency bound under the assumptions used for the trial analysis.

The estimator we will examine here is most commonly called augmented inverse propensity weighting (AIPW, which we will use throughout the paper), but it has also been referred to as double machine learning (DML) \cite{diaz-tmle-dml} or the efficient influence function (EIF) estimator \cite{rothe-rct}. This estimator is popular in the analysis of observational data and is also sometimes called the ``doubly robust'' estimator in reference to the fact that its estimates are consistent if either the propensity or outcome models are correctly specified (although other estimators also have that property). In the context of a randomized trial, the propensity score is in fact fixed and known, which means this estimator is \textit{always} consistent. The estimator is given by:

\begin{equation}
\begin{split}
\hat\tau
&= 
r\left(\hat\mu_0, \hat\mu_1\right)
\\
\hat\mu_w &= 
\Ehat{
\frac{W_{w}}{\pi_w}
    \left( Y - \hat\mu_w^{(-k)}(X) \right) +
    \hat\mu_w^{(-k)}(X)
}
\\
\hat\mu_w^{(-k)}(\cdot) &= \mathcal M \left(
\bm X_w^{(-k)}, \bm Y_w^{(-k)}
\right)
\end{split}
\label{eq:estimator}
\end{equation}

To understand the mechanics of this estimator one should work bottom-up in eq. \ref{eq:estimator}. Temporarily ignoring the $(-k)$ superscripts, the first step is to use the data and some form of machine learning algorithm $\mathcal M$ (e.g. random forest) to estimate the conditional mean functions $\hat\mu_w(X)$ in each treatment group. For this we use the data from all subjects in group $w$, denoted $(\bm X_w, \bm Y_w)$. The learned functions $\hat\mu_w(X)$ represent estimated versions of the true, unknown, conditional means $\mu_w(X) = \E{Y|X,W=w}$. With those estimates in hand, we plug our data into the middle equation in order to derive point estimates of the \textit{marginal} means $\hat\mu_w$. The right-hand side of the middle expression is interpretable as the group-specific mean outcome ``augmented'' with the model predictions in such a way that eliminates possible bias from the models. Finally, the treatment effect is estimated by plugging the estimated means into the function $r$ (top line).

Now we turn our attention to the $(-k)$ superscripts. For desirable asymptotic properties to hold without additional assumptions, it turns out that the conditional means must be \textit{cross-estimated} from the trial data \cite{Newey:2018vb, wager-stats361}. Briefly, that means we must split our trial data into $k \in {1 \dots K}$ non-overlapping folds and fit $K$ different models for $\hat\mu_w(X)$, each excluding the data from one of the folds. To notate this, let $\hat\mu_w^{-k}(X)$ denote the model trained without the $k$th fold of the data. We use the models trained \textit{without} the $k$th fold to make predictions \textit{for} the $k$th fold. In other words, when we need to get the prediction for a subject $i$, we use the model that omitted that subject from its training set. Intuitively, we do this because we do not want to ``train on the test data'' and unknowingly overfit the models. Statistically, the cross-fitting process allows us to make statements about the resulting AIPW estimator that are agnostic to the specifics of the machine learning method used to fit the model. Note that when we say ``AIPW'' throughout this article, we are always referring to AIPW with cross-fit estimates of the conditional means.

It is shown in the appendix (restating known arguments in the context of a randomized trial) that this estimator is semiparametric efficient as long as the predicted conditional means are estimated in a way that is {mean-square} consistent\footnote{
When the data are observational, the conditions are more strict because the propensity score is unknown and must be estimated. The product of the propensity and outcome model residuals must decay at a $\sqrt{n}$ rate to ensure that the resulting AIPW estimator is asymptotically normal as desired. Moreover, further assumptions are required to bound the estimated propensity scores away from 0 and 1 so that the difference of their inverses converges at a the same rate. These (rather strict) conditions are unnecessary when the treatment is randomized.
}, i.e. 

\begin{equation}
\label{eq:mse-norm}
\E{
\left(
    \hat\mu_w^{(-k)}(X) - \mu_w(X)
\right)^2
} 
\rightarrow 0
\end{equation}

This justifies the use of many popular machine learning methods $\mathcal M$ for learning the functions $\hat\mu_w(X)$ \cite{cheng1984strong, convergence-L2boost, convergence-nn, convergence-rf}.

\paragraph{Asymptotic Variance}

Recall that $r$ is the function that defines the treatment effect from the true mean outcomes $\mu_w = \E{Y_w}$. Let $r'_w = \frac{\partial r}{\partial \mu_w} (\mu_0, \mu_1)$. Any semiparametric efficient estimator is asymptotically normal
$
\sqrt{n}(\hat\tau - \tau) 
\rightsquigarrow 
N(0,\nu^{*2})
$
where $\nu^{*2}$ is the efficiency bound given by

\begin{equation}
\begin{split}
\nu^2_* &= \V{\phi}
\\
\phi &= 
    r'_0 \phi_0 +
    r'_1\phi_1
\\
\phi_w &= \frac{W_w}{\pi_w}
    \left( Y - \mu_w(X) \right) +
    (\mu_w(X) - \mu_w)
\end{split}
\label{eq:asymptotic-variance}
\end{equation}

A consistent standard error for $\hat\tau$ can be calculated from an empirical plug-in estimator $\hat\nu^2_* = \Ehat{\phi^2}$, substituting $\hat \mu_w^{(-k(i))}(X_i)$ for $\mu_w(X_i)$ and $\hat \mu_w$ for $\mu_w$ in the expression for $\phi$ \cite{wager-stats361, rosenblum-glm, rothe-rct}. Having an estimate of the asymptotic variance allows us to presume that $\hat\tau$ is distributed approximately as $N(\tau, \hat\nu^2/n)$ which allows for the construction of confidence intervals ($\hat\tau \pm 1.96 \times \hat\nu/\sqrt{n}$) and p-values ($P(|T|>\hat\tau)$ under the null $T \sim N(0, \hat\nu^2/n)$).

\subsection{Power}

Randomized trials are costly and slow, so sponsors must ensure a high chance of a statistically significant finding when the treatment truly has a clinically significant effect. This chance is assessed by a power calculation, which is performed in the design phase of the trial before any subjects have been enrolled \cite{jones-power}. The purpose of the power calculation is to help the analyst decide how many subjects would be required to ensure adequate power if the data were analyzed with a particular estimator.

Presuming that statistical significance of the result is assessed using a two-sided p-value cutoff $p < \alpha$, the probability of this event occurring when in fact the true effect is $\tau$ such that $\hat\tau \sim N(\tau, \nu^2/n)$ is

\begin{equation}
\text{Power} = 
\Phi
    \left(
        \Phi^{-1}(\alpha/2) + 
        \sqrt{n}
        \frac{\tau}{\nu}
    \right) 
+
\Phi
    \left(
        \Phi^{-1}(\alpha/2) - 
        \sqrt{n}
        \frac{\tau}{\nu}
    \right)
\label{eq:power}
\end{equation}

where $\Phi$ denotes the CDF of the standard $N(0,1)$ normal distribution. 

Assuming a particular target effect $\tau$, this formula allows the analyst to pick $n$ and an estimator $\hat\tau$ with asymptotic variance $\nu^2$ such that the resulting power is greater than a desired fraction, e.g. 80\%. The trouble in doing this is that the asymptotic variance $\nu^2$ of any given estimator ultimately depends on the specifics of the counterfactual data-generating (population) distribution $P(Y_0,Y_1,X)$ in the trial, which is unknown. For many estimators this requires estimating or guessing a large number of parameters or otherwise making strong assumptions \cite{sample-logistic, sample-logistic2, sample-logistic3, sample-poisson}.

To get around this problem, it is common to perform the power analysis with an estimator that must always attain a larger sampling variance than the one that will ultimately be used in the analysis, but which allows for tractable estimation of the asymptotic sampling variance from a small number of interpretable population-level parameters. Since the variance of the simplified estimator is greater or equal to that of the estimator used for analysis, the resulting power will be a conservative estimate of the true power (up to correct specification of the population parameters). 

For example, consider a 1:1 randomized trial with a continuous primary outcome in which the data will be analyzed with linear regression (ANCOVA). It may be shown that the difference-in-means (or ``unadjusted'') estimator $\hat\tau_{\Delta} = \Ehat{Y|W=1}-\Ehat{Y|W=0}$ has asymptotic variance $\nu^2 = 2\V{Y_0} + 2\V{Y_1}$, which is always greater than or equal to the asymptotic variance of the linear regression estimator in this case. The benefit of using the unadjusted estimator for the power calculation is that the asymptotic variance only depends on the presumed marginal variances of the two potential outcomes: $\V{Y_w}$. Under a the simplifying assumption $\V{Y_0} = \V{Y_1}$ this reduces to a single number. In any case, these are interpretable parameters that could be estimated from existing data from prior trials, registries, or electronic health records; or even guessed at by a experienced domain expert. Fixing these parameters effectively fixes $\nu$, which allows the analyst to vary $n$ until the desired power is reached. This ensures that the true power using the linear regression analysis will actually exceed the design power by some unknown amount.

\section{Approach}

The obvious downside to the usual conservative approach to powering trials is that more subjects are enrolled than are actually necessary to attain the design power (assuming correct specification of the required population parameters). This is of no scientific concern since we may be confident that the analysis is well-powered, but it is of enormous practical concern because each additional subject adds significant cost, time, and complexity to the trial.

\subsection{Accounting for Semiparametric Efficiency}

Our aim here is therefore to examine the asymptotic variance of the semiparametric estimators described in section \ref{sec:estimator} and reduce it down to a small number of population-level parameters which are estimable from data and/or have natural interpretations. An expression of that kind would allow the analyst to proceed with the power calculation without substituting the anticipated (larger) variance from a simplified estimator. This, in turn, would remove undesired conservatism and allow for smaller trials that maintain their design power.

\begin{theorem}
Let $\sigma_w^2 \equiv \V{Y_w}$ be the marginal outcome variances in each treatment arm and let $\kappa^2_w \equiv \E{\V{Y_w|X}}$ be the corresponding average conditional variances. Moreover define $\gamma = \text{Corr}[\mu_0(X), \mu_1(X)]$, the correlation between the conditional means and recall $r'_w = \frac{\partial r}{\partial \mu_w} (\mu_0, \mu_1)$.

Subject to mild regularity conditions on the data-generating distribution, the asymptotic variance of any semiparametric efficient estimator of the parameter $\tau = r(\mu_0, \mu_1)$ is 

\begin{equation}
\begin{split}
    \nu_*^2 
    & = 
        r_0'^2 \left(
            \frac{\pi_1}{\pi_0} \kappa^2_0 + \sigma_0^2
        \right) +
        r_1'^2 \left(
            \frac{\pi_0}{\pi_1} \kappa^2_1 + \sigma_1^2
        \right) - 
        2 |r_0'r_1'| \gamma \sqrt{(\sigma_0^2 - \kappa^2_0)(\sigma_1^2 - \kappa^2_1)}
\end{split}
\label{eq:asymptotic-variance-parametrized}
\end{equation} 

\label{thm:asymptotic-variance}
\end{theorem}

The proof of this, which requires nothing but algebra, is given in the appendix.

\paragraph{Special cases} There are a few interesting special cases of this relationship. The first is when $\mu_w(X) = \mu_w$ are constants. In this case we have $\kappa^2_w = \sigma_w^2$ (note $\kappa^2_w \leq \sigma_w^2$ by the law of total variance) and the above reduces to 
$
\nu^2_* = 
 r_0'^2 \frac{\sigma_0^2}{\pi_0}
 +
 r_1'^2 \frac{\sigma_1^2}{\pi_1} 
$, the variance of the unadjusted (difference in means) estimator. In other words, the unadjusted estimator is efficient when the conditional means are constant, as should be obvious because the covariates impart no exploitable information.

A special case that illuminates the role of $\gamma$ is when $\pi_0=\pi_1$, $\sigma_0 = \sigma_1 = \sigma$, and $\kappa^2_0 = \kappa^2_1 = \kappa^2$. Presume the estimand of interest is $\tau = \mu_1 - \mu_0$ such that $r_0'=-1$ and $r_0'=1$. In this case the above reduces to $\nu^2_* = 2[(1-\gamma)\sigma^2 + (1+\gamma)\kappa^2]$. Compare this to the asymptotic variance of the unadjusted estimator under these conditions, which is $\nu_\Delta^2 = 4\sigma^2$. Indeed, in the worst case scenario (when $\gamma = -1$) our semiparametric efficient estimator gives the same asymptotic variance as the unadjusted estimator, meaning that no efficiency gain is possible from covariate adjustment. In the best-case scenario ($\gamma =1$) the asymptotic variance is $\nu^2_* = 4\kappa^2$. Note that this case obtains when there is a constant treatment effect across the population. When $\gamma = 0$ we get the intermediate $\nu^2_* = 2\sigma^2 + 2\kappa^2$. 

The takeaway is that $\gamma$ mediates the the extent to which the asymptotic variance depends on the \textit{marginal variance} (worst case) vs. the \textit{average conditional variance} (best case) of the potential outcomes in each treatment arm. This again should be intuitive. The law of total variance states that the average conditional variance is the component of the marginal variance that is left over when the variance of the conditional mean is subtracted off. By performing a nonlinear adjustment for the covariates, we come close to recovering the true conditional mean in large samples and we can ``explain away'' exactly that component of the variance.

\subsection{Prospective estimation}

Recall that our goal is to estimate the asymptotic variance \textit{before} running the trial so that we may use it in conjunction with eq. \ref{eq:power} to find an sample size that should suffice to attain a desired level of power. The relationship in thm. \ref{thm:asymptotic-variance} shows that only $\sigma_w^2$, $\kappa^2_w$ and $\gamma$ are required to calculate $\nu_*$ (after specifying $\mu_w$ in order to set the target effect $\tau$, which fixes $r_w'$ as well). Our task is therefore to hypothesize values or bounds for these parameters or to estimate them using pre-existing data.

The marginal variances $\sigma^2_w$ should be familiar to most analysts. Usually there is existing knowledge about the marginal variance under standard-of-care from registry data, electronic health records, or prior studies on similar populations. This variance is taken to be $\sigma^2_0$ and it is most often assumed that $\sigma_1 = \sigma_0$ because there is rarely reliable data on treatment-arm outcomes (since the treatment is often new and experimental {and phase I/II trials are typically small}). {If treatment-arm data is available, then $\sigma_1$ can be estimated independently.}

The average conditional variances $\kappa^2_w$ are less-familiar quantities, but they also admit natural interpretations and estimators. One way to estimate an upper bound on the average conditional variance in either treatment arm is to average known (marginal) variances across sub-populations defined by the planned adjustment covariates. For instance, presume the anticipated trial population consists of about equal numbers of men and women and that biological sex is a planned adjustment covariate. Then the mean of the marginal outcome variance among men in treatment arm $w$ and the marginal outcome variance among women in treatment arm $w$ is a consistent estimator for an upper bound on the average conditional variance in treatment arm $w$. The population can be arbitrarily divided in this way as many times as existing data permits as long as the splits are pre-specified. By the bounding argument, this estimate of average conditional variance is likely to be larger than the true value, yielding a conservative estimate for power that still offers gains over the traditional difference-in-means power calculation. {If treatment-arm data are not available or are too scant to allow for subdivision, one might assume that $\kappa^2_1 = \kappa^2_0$, or, very conservatively, that $\kappa^2_1 = \sigma_1^2$ (recall $\kappa^2_w \leq \sigma_w^2$).} 

An alternative and more data-intensive method for estimating $\kappa^2_w$ is to use a machine learning model in combination with historical data. Note that $\kappa^2_w$ is in fact the Bayes mean-squared error (MSE) for the prediction problem of estimating $\E{Y_w|X}$.\footnote{
By the law of total expectation, 
$\E{(Y_w-\mu_w(X))^2} = \E{\E{(Y_w-\mu_w(X))^2|X}}$. 
The internal expectation is the definition of the conditional variance, i.e.
$\E{(Y_w-\mu_w(X))^2|X} \equiv \V{Y_w|X}$.
}  So if there are existing data for $(X, Y_w)$, the test-set MSE for a machine learning model trained using those data is a consistent estimator for an upper bound on the Bayes MSE because the Bayes MSE is by definition the MSE of the best possible model.\footnote{Note that it is always the \textit{MSE} that is the quantity of interest, even if the outcome is binary, etc. In the binary outcome case the model's prediction (an estimate of $\E{Y_w|X}$) represents the probability of the outcome. MSE in this case is equivalent to the ``Brier score''.} This process would also produce a usable upper bound even if only a subset of the planned covariates were available in the historical data because $\V{Y|X,Z} \le \V{Y|X}$.

Like $\sigma_1^2$ and $\kappa^2_1$, the value $\gamma$ depends on the behavior of the treatment arm and so it is not estimable using historical control-arm or standard-of-care data alone. However, there are interpretable domain assumptions that bound $\gamma$. For example, in many cases it is assumed that the effect of treatment is additively constant across the population (i.e. $\mu_1(X) = \mu_0(X) + \tau$). When that is the case, $\gamma = 1$. It is unreasonable to assume this relationship would obtain exactly, but we recommend $\gamma \ge 0$ as a conservative lower bound that provides substantial wiggle room for treatment effect heterogeneity. Indeed, Cauchy-Schwarz and a little algebra\footnote{
Abbreviate 
$\V{\mu_0} = \V{\mu_0(X)}$ and 
$\V{\tau} = \V{\mu_1(X)-\mu_0(X)}$.
$
\V{\mu_0} \ge \V{\tau} 
\iff \V{\mu_0} \ge \sqrt{\V{\tau}\V{\mu_0}}
\implies \V{\mu_0} \ge \left|\lambda \sqrt{\V{\tau}\V{\mu_0}}\right| \ \forall \lambda \in [-1,1]
\implies \V{\mu_0} \ge \left| \C{\tau, \mu_0} \right|
\implies \V{\mu_0} + \C{\tau, \mu_0} \ge 0
\iff \C{\mu_0 + \tau, \mu_0} \ge 0
\iff \C{\mu_1, \mu_0} \ge 0
\iff \gamma \ge 0
$. Replace $\mu_0(X)$ with $\mu_1(X)$ and proceed in the same way to show the equivalent for $\mu_1(X)$.
} 
show that $\V{\mu_0(X)} \ge \V{\mu_1(X) - \mu_0(X)}$ or $\V{\mu_1(X)} \ge \V{\mu_1(X) - \mu_0(X)}$ are sufficient conditions to ensure $\gamma \ge 0$. So as long as the heterogeneity of (additive) effect is less than the heterogeneity in expected outcomes in either treatment arm, $\gamma \ge 0$ will certainly hold. {In the case where a substantial amount of treatment arm data is available, it may be possible to estimate $\gamma$ by calculating the empirical correlation coefficient between the out-of-sample predictions of estimated treatment-arm and control-arm models. At a minimum, an estimate of this kind could be used to confirm $\gamma \ge 0$, which would justify the conservative assumption that $\gamma=0$}.

\section{Demonstration}

{
Here we demonstrate via simulations and a case study that our approach for prospective sample size estimation results in trials that attain their design power with fewer subjects that would be required with existing methods. We focus on the use-case where there are no available treatment arm data before the trial is run because these data are usually scant at best (e.g. come from a relatively small phase II trial) and likely insufficient to reliably estimate $\kappa_1$ and $\gamma$. On the other hand, standard-of-care data are often plentiful and can be used as a reasonable proxy for control-arm data. In practice we recommend using any existing treatment-arm data to verify that $\sigma_0 \gtrapprox \sigma_1$, $\kappa_0 \gtrapprox \kappa_1$, and $\gamma \ge 0$ so that power estimates are ensured to be conservative.
}

\subsection{Simulation}

\paragraph{Methods}
Our treatment effect of interest was the difference-in-means of a continuous outcome (thus $r_0'=-1$, $r_1'=1$). In each simulation scenario we fixed a counterfactual data-generating process $P(Y_1, Y_0, X)$. We sampled a dataset of 10,000 observations of $X, Y_0$, which represented a historical sample (e.g. prior trial control arms, registry data) that we used to estimate the population parameters we needed for the sample size calculation. Our estimates of these parameters were:
\begin{equation}
\begin{split}
\hat\sigma_0 &= \hat\sigma_1 = \Vhat{Y_0} \\
\hat\kappa_0 &= \hat\kappa_1 = \text{cross validation RMSE of an ensemble model trained on $\bm X, \bm Y_0$} \\
\hat\gamma &= 0
\label{eq:power-params}
\end{split}
\end{equation}

The ensemble model we used was a cross-validated selection between linear regression (LR),  5-nearest-neighbors (5-kNN), {and gradient boosted trees (GBM; 50 depth-5 trees, all other parameters left at their scikit-learn defaults \cite{scikit-learn})}. {Note that kNN explicitly satisfies the {mean-square}  consistency (eq. \ref{eq:mse-norm}) required for AIPW to be semiparametric efficient \cite{cheng1984strong}}. Nested cross-validation to select between these three models was used within the outer cross validation loop used to estimate RMSE.

We used these parameter estimates to estimate the asymptotic variances of the AIPW and unadjusted estimators in a 1:1 randomized trial ($\pi_w=1/2$) according to the formulae

\begin{equation}
\begin{split}
    \hat\nu^2_{\text{AIPW}} &= 
        r_0'^2 \left(
            \frac{\pi_1}{\pi_0} \hat\kappa^2_0 + \hat\sigma_0^2
        \right) +
        r_1'^2 \left(
            \frac{\pi_0}{\pi_1} \hat\kappa^2_1 + \hat\sigma_1^2
        \right) - 
        2 |r_0'r_1'| \gamma \sqrt{(\hat\sigma_0^2 - \hat\kappa^2_0)(\hat\sigma_1^2 - \hat\kappa^2_1)} 
    \\
    \hat\nu^2_{\text{unadj}} &= 
     r_0'^2 \frac{\hat\sigma_0^2}{\pi_0}
     +
     r_1'^2 \frac{\hat\sigma_1^2}{\pi_1} 
\label{eq:estimated-variances}
\end{split}
\end{equation}

We also calculated the \textit{true} asymptotic variance bound $\nu_*^2$ (thm \ref{thm:asymptotic-variance}) by extracting the true values of $\sigma_w$, $\kappa_w$, and $\gamma$ from the counterfactual distribution. {Note that there is no ``$\hat\nu^2_\text{ANCOVA}$'' because there does not exist a parsimonious asymptotic variance formula for ANCOVA that involves only a small number of population parameters \cite{procova}. Trials analyzed with ANCOVA typically assume unadjusted estimation for sample size calculation, which is conservative.}

We used the prospective variance estimates $\hat\nu_{\text{AIPW}}^2$ and $\hat\nu_{\text{unadj}}^2$ to choose sample sizes $n^\dagger_{\text{AIPW}}$, $n^\dagger_{\text{unadj}}$ for a hypothesized trial according to the formula 
\begin{equation}
n^\dagger =
\begin{cases}
& \text{argmin}_n \  n \\
& \text{s.t.} \ 1-\beta < \Phi
    \left(
        \Phi^{-1}(\alpha/2) + 
        \sqrt{n}
        \frac{\tau}{\nu}
    \right) 
+
\Phi
    \left(
        \Phi^{-1}(\alpha/2) - 
        \sqrt{n}
        \frac{\tau}{\nu}
\right)
\end{cases}
\label{eq:find-n}
\end{equation}

plugging in $\hat\nu_{\text{AIPW}}$ and $\hat\nu_{\text{unadj}}$ for $\nu$, respectively. $n^\dagger_{\text{unadj}}$ represents the enrollment of the trial that we would have calculated if we used the standard, unadjusted power formula, which we know to be overly conservative. $n^\dagger_{\text{AIPW}}$ represents the enrollment of the trial that we would have calculated if we used our proposed power formula implied from theorem \ref{thm:asymptotic-variance} and specified that the analysis of the trial would be conducted with an AIPW estimator. We also calculated the enrollment using $\nu_*^2$ to obtain a value $n_*^\dagger$, which represents the trial size we would need to attain design power if we were to use an ``oracle'' AIPW estimator with direct access to the true functions $\mu_w(X)$. It would not be possible to do this in practice, but it is useful to study the behavior of the oracle estimator because it represents an effective upper bound on performance.

The desired power level $1-\beta$ was set to 0.8 and the significance level $\alpha$ was set to 0.05 in all experiments. The target treatment effect $\tau$ was the true effect from the counterfactual distribution.

To calculate empirical power we then simulated 1:1 randomized trials of a variety of sizes ($n$). In each simulated trial, we drew data from the specified counterfactual distribution, randomly assigned treatment, and hid the unobserved potential outcomes. We then used an ``oracle'' AIPW estimator, a cross-fit AIPW estimator (using cross-validated selection between GBM, 5-kNN, and LR to learn $\hat\mu^{(-k)}(\cdot)$), a standard main-terms ANCOVA estimator with robust standard errors (HC0, \cite{long-hc3}), and an unadjusted estimator to estimate the treatment effect and a standard error from the data. We recorded if the result was statistically significant in each case. We repeated each experiment 1000 times and calculated the power of each estimator by computing the fraction of significant results. This was repeated for each trial size $n$.

The objective of these simulations was to see if the power of the AIPW estimator for $n>n^\dagger_{\text{AIPW}}$ indeed exceeded the design power $1-\beta = 0.8$. Secondarily, we were also interested in the amount by which  $n^\dagger_* < n^\dagger_{\text{AIPW}} < n^\dagger_{\text{unadj}}$, the overall amount by which AIPW increased power over the ANCOVA or unadjusted anayses, the amount of conservatism in $n^\dagger_{\text{AIPW}}$, and the difference in performance between the cross-fit and oracle AIPW estimators.

We also conducted simulations in which the average treatment effect was 0, in which case we interpreted the rate of significant results as the empirical type I error of each estimator. The purpose of those simulations was to confirm that all of our estimators control type I error in large samples.

{

Lastly, we repeated each of our experiments with the AIPW estimator using one of GBM, kNN, or LR by itself as the estimator of the conditional means. For these experiments we used the same hyperparameter settings for each learner as were used in the ensemble learner. We used the same methods as above to calculate prospective sample sizes for each of these designs. The purpose of these experiments was to assess how sensitive the AIPW estimator is to the choice of learner and, more importantly, to see whether the corresponding prospectively-calculated sample sizes (i.e. $n^\dagger_\text{AIPW(kNN)}$, $n^\dagger_\text{AIPW(GBM)}$, and  $n^\dagger_\text{AIPW(LR)}$) resulted in well-powered trials in all cases.
}

\paragraph{Scenarios}
Each simulation scenario was fully specified by the choice of counterfactual distribution $P(Y_1, Y_0, X) = P(Y_1|X)P(Y_0|X)P(X)$. We chose the scenarios to reflect our real-world presumptions that $\sigma_0 \gtrapprox \sigma_1$ as well as $\kappa_0 \approx \kappa_1$ and $\gamma \ge 0$. In other words, treatment did not substantially affect the heterogeneity in outcomes and any heterogeneity in treatment effect was on the order of or smaller than the heterogeneity in expected outcomes.\footnote{
Without these conditions, it is not possible to conservatively use historical control data to estimate the necessary parameters for either our power formula or the unadjusted power formula.
}
For convenience of plotting, we specified the distributions so that $n^\dagger_{\text{AIPW}}$ and $n^\dagger_{\text{unadj}}$ would come out as numbers on the order of $10^2$.

Within these constraints we explored four different scenarios including linear and nonlinear functions for the conditional means and the presence or absence of treatment effect heterogeneity. While no set of simulations can possibly come close to spanning the full space of possible distributions, we chose these to be illustrative of important archetypal scenarios one might find in a randomized trial.

Our simulations follow the format of \citet{procova}. In all cases, the distribution of covariates $P(X)$ was a 10-dimensional uniform random variable in the prism $[-1,1]^{10}$. $P(Y_w|X)$ were of a Gaussian quadratic-mean form $\mathcal N(aX^\top \mathds 1 X + bX^\top \mathds 1 + c, 1)$ in all scenarios ($\mathds 1$ is a matrix or vector of 1s with appropriate shape implied). The parameter $a$ controls the degree of non-linearity in this specification, with $a=0$ representing the linear case. Treatment effect heterogeneity refers to the situation in which $a$ or $b$ is different for $P(Y_0|X)$, and $P(Y_1|X)$. The specific values of $a$, $b$, and $c$ for each scenario are shown in Table \ref{tbl:sim_scenarios}.

\begin{table}[h!]
\centering
\begin{tabular}{|c||c|c|c|c|c|c|}
\hline
\multirow{2}{*}{Scenario} &
\multicolumn{3}{c|}{$P(Y_0|X)$} &
\multicolumn{3}{c|}{$P(Y_1|X)$}
\\ \cline{2-7}
 &
$a_0$ & $b_0$ & $c_0$ &
$a_1$ & $b_1$ & $c_1$
\\ \hline \hline
{linear, constant} &
0 &  1 &  0 &
0 &  1 &  1/2
\\ 
{linear, heterogeneous} &
0 &  1 &  0 &
0 &  0 &  1/2
\\
{nonlinear, constant} &
1 &  1 &  0 &
1 &  1 &  1
\\ 
{nonlinear, heterogeneous} &
1 &  1 &  0 &
1 &  0 &  1
\\ \hline
\end{tabular}
\caption{Parameters for all simulation scenarios.}
\label{tbl:sim_scenarios}
\end{table}

The scenarios and parameters for the type I error simulations were identical except that $c_0$ was modified in each scenario such that the average treatment effect became 0.

\paragraph{Results}

\begin{figure}[h]
    \centering
    \includegraphics[width=1\textwidth]{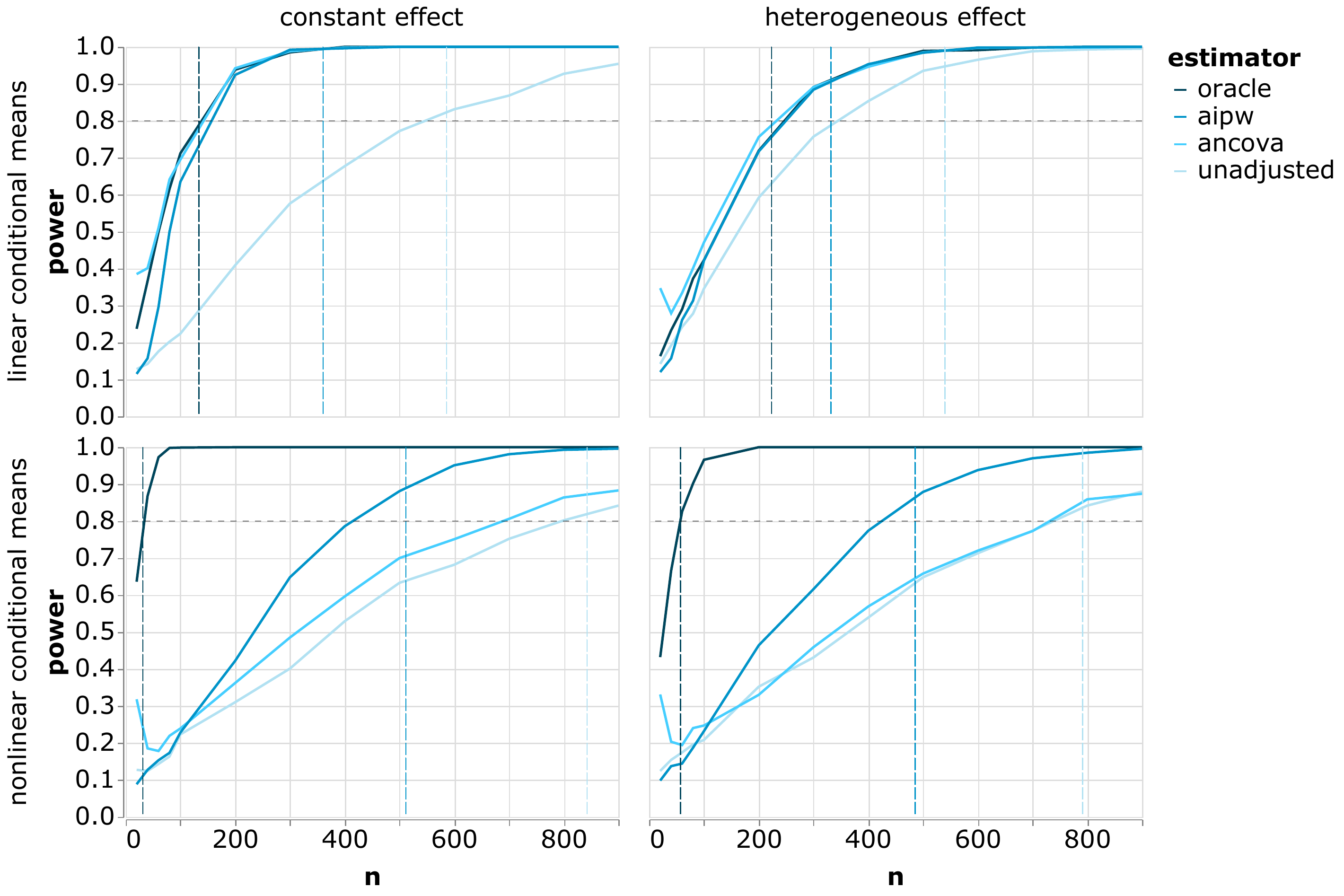}
    \caption{Empirical power and prospectively-calculated enrollment targets. The dotted vertical lines correspond to enrollment targets, which were prospectively calculated using parameters estimated from a ``historical'' dataset. The lightest blue corresponds to the enrollment required for 80\% power as dictated by the unadjusted estimator, darker blue corresponds to the AIPW estimator, and darkest to the oracle AIPW estimator. The curves indicate the true empirical power of each estimator (including ANCOVA for comparison) as the enrollment is varied. The horizontal dotted line represents the $80\%$ target design power. {A cross-validated ensemble of GBM, 5-kNN, and linear regression was used for the AIPW estimator and to prospectively calculate the corresponding sample size.}}
    \label{fig:power}
\end{figure}

Our results (figure \ref{fig:power}) show that trials analyzed with the AIPW estimator attain power greater than 80\% at the corresponding enrollment targets. Since these targets were prospectively calculated, this provides empirical evidence that our methods can be applied in the design phase to power trials as long as the analysis plan specifies an AIPW (or other semiparametric efficient) estimator. Moreover the enrollment target is still likely to be conservative by some amount and robust to heterogeneity of the treatment effect and nonlinear conditional means. 

The enrollment target calculated by assuming the unadjusted estimator is also robust to different conditions but is far more conservative than is necessary when any kind of covariate adjustment is employed in the analysis. Indeed, we observe potential sample size savings of $\sim 35\%$ in these simulations when using the AIPW-based enrollment target. However, one advantage of the unadjusted enrollment target is the trial will attain or exceed its design power regardless of what estimator is specified in the analysis plan, as long as it is as efficient or more efficient than the unadjusted estimator.

We also observe near-perfect agreement between the power predicted at the oracle enrollment target and the empirical power attained by the oracle estimator at that sample size. This is strong empirical evidence for the validity of theorem \ref{thm:asymptotic-variance}. More importantly, however, the discrepancy between the oracle enrollment target and the estimated AIPW enrollment target suggests that it is not necessarily beneficial to have access to the true population parameters ($\sigma_w$, $\kappa_w$, $\gamma$) at design time because the AIPW estimator is typically far from its optimal efficiency in practice. Trials analyzed with a real-world (not oracle) AIPW estimator will usually \textit{not} attain their design power at the \textit{oracle} enrollment target. In other words, there is benefit to having to \textit{estimate} the population parameters from data because using those estimates in the sample size calculation better reflects the finite-sample performance of the AIPW estimator.

Our results also confirm what is known about the relative efficiency of AIPW, ANCOVA, and unadjusted estimation. Main-terms ANCOVA performs as well as the oracle in the scenario with linear conditional means and a constant effect because it is perfectly specified in that setting. Our AIPW estimator is also near-optimal in that case because it includes a linear regression in its ensemble model of the conditional means. On the other hand, AIPW greatly exceeds ANCOVA in power when there is any nonlinearity. {The efficiency gain over unadjusted estimation is relatively lessened here, likely because our kNN estimation of the conditional mean functions converges much more slowly than the linear regression does in the linear case.} In these settings, there is also a large difference between the performance of the oracle and real-world AIPW estimators. This suggests that there is a substantial opportunity to improve the quality of the conditional mean modeling in the AIPW estimator (e.g. with stronger machine learning methods or via models pre-trained on large external datasets and fine-tuned on the trial data). 

\begin{figure}[h!]
    \centering
    \includegraphics[width=1\textwidth]{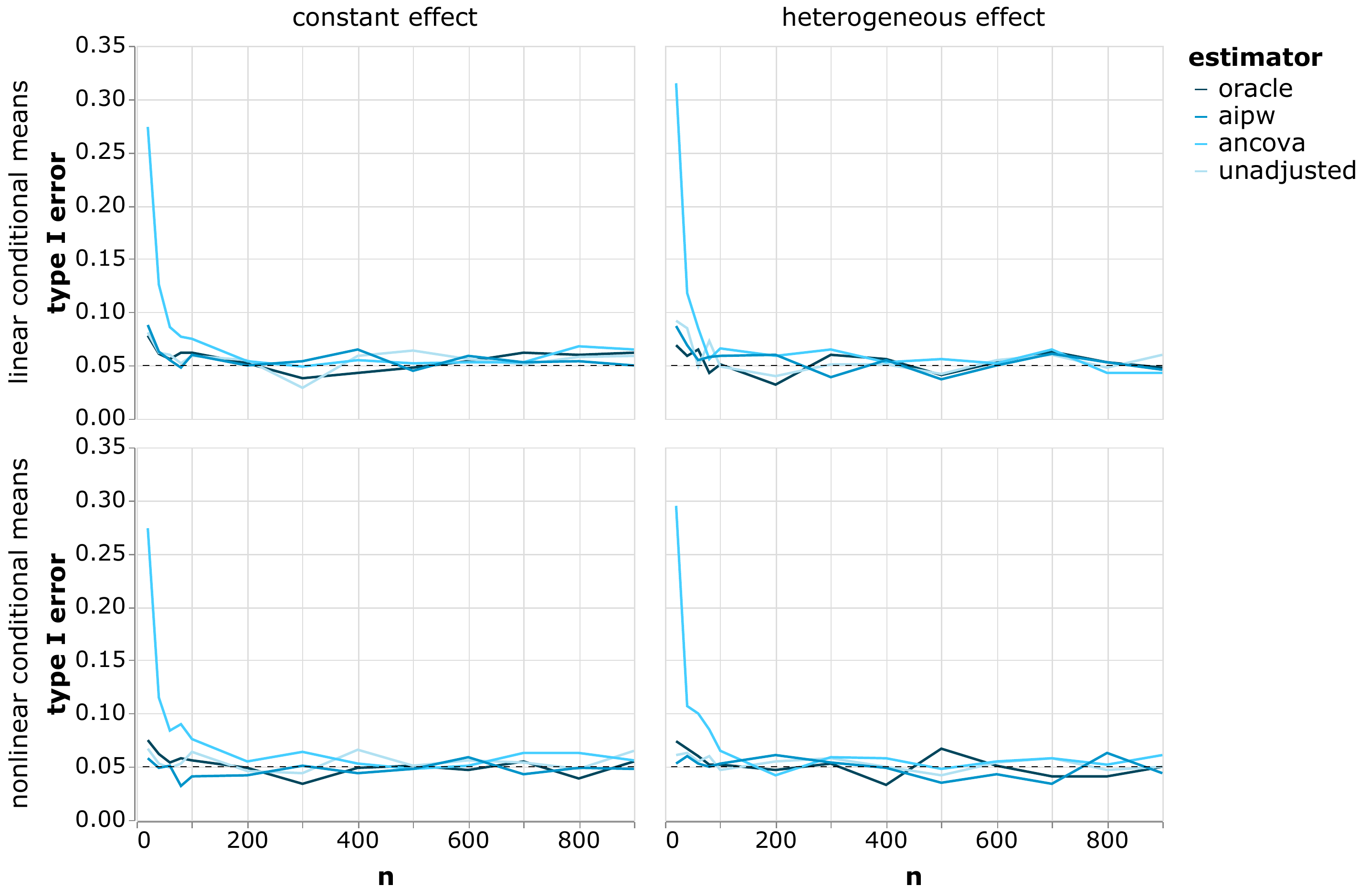}
    \caption{Empirical type I error of the orracle, AIPW, ANCOVA, and unadjusted estimators in our simulations. The horizontal dotted line represents the $5\%$ target type I error rate.}
    \label{fig:error}
\end{figure}

The results of our simulations in scenarios with no treatment effect (figure \ref{fig:error}) confirm that the AIPW estimator controls type I error in large samples \cite{zivich}. As expected due to the asymptotic nature of the inference, we observed some inflation of type I error for all estimators in very small samples ($n \lesssim 100$).\footnote{The small-sample behavior of ANCOVA with robust standard errors has been the subject of extensive prior work \cite{long-hc3}.}

\begin{figure}[h]
    \centering
    \includegraphics[width=0.6\textwidth]{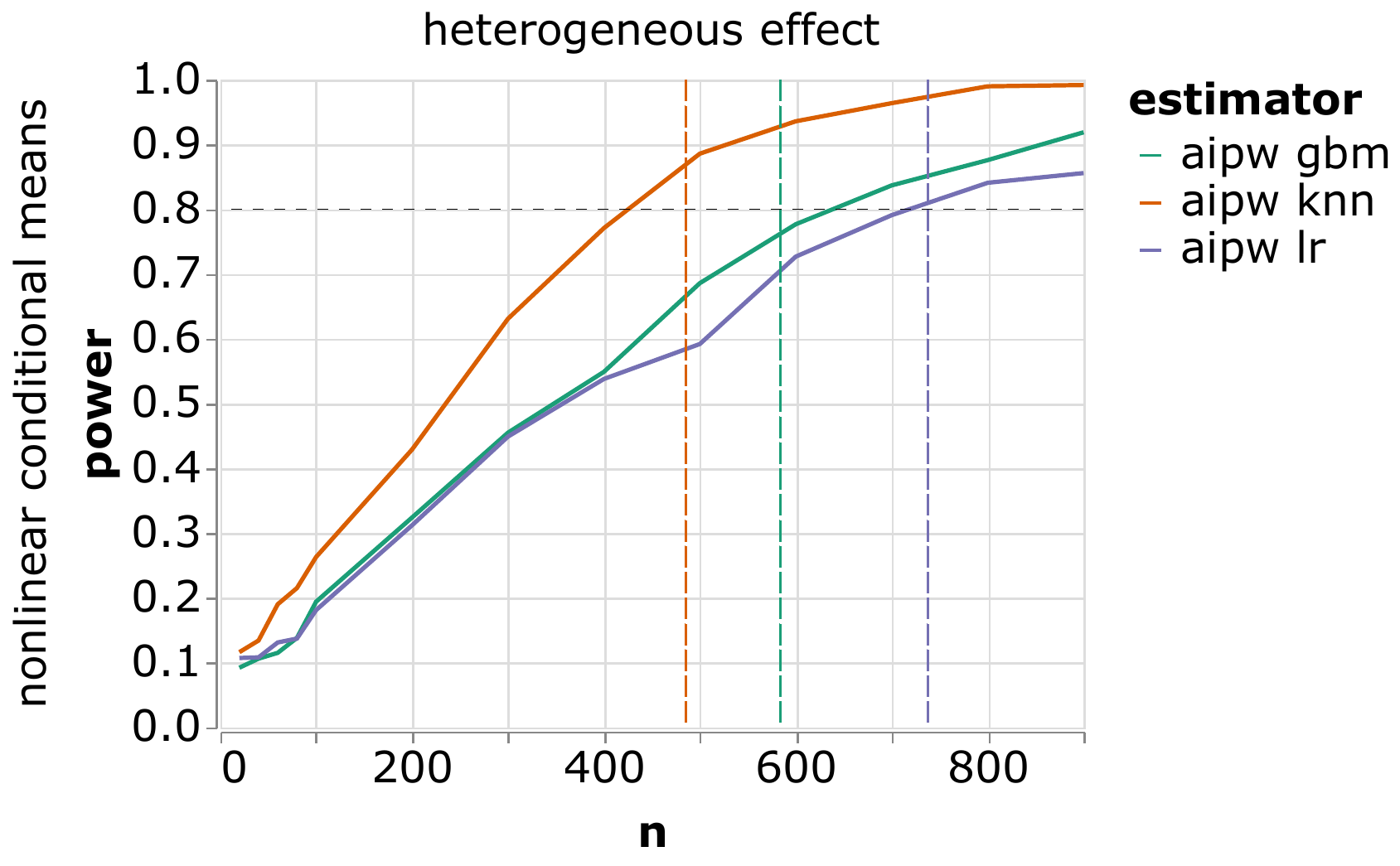}
    \caption{
    {
    Empirical power and prospectively-calculated enrollment targets for AIPW estimators with different learners used to estimate the conditional means. Colors differentiate the learners (gbm: gradient boosted trees, knn: k nearest neighbors, lr: linear regression) with hyperparameters as described above. Other visual elements are as in figure \ref{fig:power}. Additional simulation scenarios are shown in the appendix.
    }
    }
    \label{fig:power-M}
\end{figure}

{
Our experiment varying the learner used to estimate the conditional means for the AIPW estimator shows that our procedure is largely robust to what machine learning method is chosen (figure \ref{fig:power-M}). However, the experiments with GBM produced trials that in some cases were slightly under-powered, which could be of concern. Because of this, we recommend using a diverse cross-validated (or stacked) ensemble of learning algorithms with varied tuning parameters.

Surprisingly, prospective power estimation worked well when linear regression was used to estimate the conditional means for AIPW. In this case the {mean-square}  consistency condition (eq. \ref{eq:mse-norm}) is clearly not satisfied so theorem \ref{thm:asymptotic-variance} may not hold; thus it may not be prudent to rely on this empirical result. Using linear regression in the AIPW estimator gave less power than other learners when there was any nonlinearity in the conditional means. Note that including linear regression in an ensemble model ensemble effectively guarantees that AIPW will exceed the power of ANCOVA. 
}

\subsection{Case Study}

In addition to the simulations presented above, we re-analyzed data from an existing trial to demonstrate how our semiparametric efficient estimation can be taken into account when prospectively powering trials. Much of the subsequent description is shared with \citet{procova}.

\paragraph{Trial}

Our demonstration trial, reported by \citet{dha}, was conducted to determine if docosahexaenoic acid (DHA) supplementation slows cognitive and functional decline for individuals with mild to moderate Alzheimer's disease. The trial was performed through the Alzheimer's Disease Cooperative Study (ADCS), a consortium of academic medical centers and private Alzheimer disease clinics funded by the National Institute on Aging to conduct clinical trials on Alzheimer disease. 

\citet{dha} randomized 238 subjects to a treatment arm given DHA and 164 subjects to a control arm given placebo. This trial measured a number of covariates at baseline including demographics and patient characteristics (e.g. sex, age, region, weight), lab tests (e.g. blood pressure, ApoE4 status \cite{apoe4}), and component scores of cognitive tests. A full list of the 37 covariates we used is reported in \citet{procova}. The primary outcome of interest for our reanalysis was the increase in the Alzheimer's Disease Assessment Scale - Cognitive Subscale (ADAS-Cog 11, a quantitative measure of cognitive ability) \cite{adas-cog} over the duration of the trial (18 months). 

\paragraph{Methods}
Before examining the trial data, we estimated the population parameters required for a sample size calculation from a historical dataset comprised of 6,919 early-stage Alzheimer's patients. These data came from the Alzheimer's Disease Neuroimaging Initiative (ADNI) and the Critical Path for Alzheimer's Disease (CPAD) database \cite{cpad1,cpad2}, and included measurements of ADAS-Cog 11 at 6-month, or more frequent, intervals post-baseline. The ADNI dataset is made up of longitudinal data from 4 sequential large observational studies in Alzheimer's disease, while the CPAD dataset is made up of control arm data from 29 Alzheimer's disease randomized trials. These data also included the same baseline covariates as were measured in the DHA trial (imputed to a column mean where missing). 

Once these data were prepared, we used the procedure detailed in eq. \ref{eq:power-params} to estimate values for $\sigma_w$, $\kappa_w$, and $\gamma$. We set $\pi_w$ to the observed treatment ratios in the trial. We then used these values to calculate the sample sizes that would be required to detect a clinically meaningful target treatment effect of $2.7$ points with power $>80\%$ given a significance level of $0.05$, assuming either an unadjusted analysis ($n^\dagger_{\text{unadj}}$) or an efficient analysis ($n^\dagger_{\text{AIPW}}$) according to the procedure described in eqs. \ref{eq:estimated-variances} and \ref{eq:find-n}. This was done ``prospectively'', i.e. before we performed any analysis of the trial data.

We then sought to emulate what \textit{would} have happened had we executed the trial with either a) the smaller enrollment ($n^\dagger_{\text{AIPW}}$) and AIPW in the analysis plan or b) the larger enrollment ($n^\dagger_{\text{unadj}}$) and unadjusted estimation in the analysis plan. We consider these to be two different ``trial designs''. As in our simulations, we pre-specified the AIPW estimator to use a cross-validated selection between 5-kNN, GBM, and linear regression in the estimation of conditional means.

The idea of this case study was to see whether the final inference about the treatment effect in both designs turned out to be similar, despite the different number of subjects. To emulate running these two trials side-by-side, in each case we took a bootstrap sample of $n^\dagger$ subjects with observed outcomes from the full trial dataset (in the original randomization ratio) and then analyzed that sample of data with the specified estimator. We repeated this process 500 times for each design to average out the resampling variability. We report the mean estimated treatment effect and the mean estimated variance from executing each trial design.

\paragraph{Results}

Our sample size calculation assuming unadjusted estimation yielded a required enrollment of $n^\dagger_{\text{unadj}} = 272$ (115 controls, 157 treated). Assuming an efficient estimator we obtained $n^\dagger_{\text{AIPW}} = 243$ (103 controls, 140 treated). Sample size calculation leveraging efficient estimation therefore yielded a sample size savings of $\sim 10\%$ in this example.

The mean effect estimate in the unadjusted design was -0.140 with a mean variance of 1.047. The mean effect estimate in the AIPW design was 0.002 with a mean variance of 1.059. The two estimates are qualitatively identical (null effect). Moreover, the variance of both estimates are nearly identical despite the fact that the AIPW design used 10\% fewer subjects. Although we cannot estimate empirical power in this case because we do not know the true treatment effect and cannot repeat the trial, the similarity of the estimated variances implies that both designs are likely to have about equal power.

\section{Discussion}

Our theory, simulations, and case study all demonstrate how efficient estimation can be exploited to design smaller trials that attain their design power without sacrificing type I error control. The required sample size calculation is parsimonious and requires the analyst to provide only a small number of population parameters. Under mild assumptions, these parameters are all estimable from historical data with off-the-shelf techniques. The resulting sample size is typically much smaller than would otherwise be required and often still provides a margin of safety. This margin can always be increased by conservative estimation of the required population parameters.

Although our simulations and case study focused on the case of a continuous outcome, our theory shows that theorem $\ref{thm:asymptotic-variance}$ applies equally well in, e.g., a trial with a binary outcome and an odds ratio estimand. {Moreover, although our simulations and case study focused only on AIPW, our theory holds for any semiparametric efficient estimator in the context of randomized trial data (e.g. TMLE).}

Also, for simplicity of presentation, we avoided discussion of { missing covariate data, }dropout, inter-current events, and other complicating factors in trial design. {For example, consider a trial where the outcome is accurately measured, but less attention is paid to the covariates. Any missing covariate values would need to be imputed (e.g. using column means) before adjusted analysis. Imputation of covariates does not affect type I error control, but the average conditional variances $\kappa_w$ estimated from historical data with less missingness may be too small in this case and the power may be overestimated. These phenomena were not modeled in our simulations or case study. Indeed, there are no general-purpose solutions to problems like these, many of which also affect standard trial designs. In practice, statisticians should use heuristic methods (e.g. artificially censoring historical covariate data) and/or conservative assumptions (e.g. multipliers on estimated parameters) as necessary to counteract the effects of these imperfections on sample size estimation}.

In addition to our power formula, we contribute further evidence that cross-fit AIPW in particular is a safe and effective estimator in the context of randomized trials. Our results show that it can easily meet or exceed the power of ANCOVA and even appears to have better control of type I error in small samples. Although AIPW makes use of innovations in ``black-box'' machine learning, it is no harder to interpret the estimated treatment effect. {If anything, inference becomes clearer because it provides more conclusive evidence. This benefit comes without any meaningful cost because AIPW estimation requires no additional assumptions to ensure type I error control in randomized trials.}

Moreover, AIPW with black-box machine learning is still perfectly compatible with pre-specification. In order to pre-specify an analysis with AIPW, one must clearly lay out the set of learners (e.g. kNN, linear regression) and hyperparameters (e.g. number of neighbors, regularization strength) that will be considered and how their predictions will be selected among, combined, etc. (e.g. selection via cross-validation, super-learning \cite{esl:2009wc, vdl-superlearner}). All of these choices can and should be made at design time, as we demonstrated in our case study. {The only requirement for theorem \ref{thm:asymptotic-variance} to hold for the AIPW estimator is that the learner (or ensemble) be {mean-square}  consistent in estimating the conditional mean functions. As such, it is generally not appropriate to use (only) linear models for this purpose.} However, the analyst can rest assured that the AIPW estimator will control type I error in large samples regardless of the learner as long as it is pre-specified. {Moreover, our empirical evidence suggests that trials may be well-powered even if the assumptions of theorem \ref{thm:asymptotic-variance} are violated.}

Our results also suggest that further sample size reductions are possible. In cases where the conditional mean functions were nonlinear, the oracle AIPW estimator outperformed the real-world AIPW estimator by a large margin. The only difference between these two estimators is that the true conditional means are known to the oracle instead of estimated. This implies that better conditional mean models could yield large improvements in power or be translated to larger sample size reductions. Our experiments used an ensemble of 5-kNN, GBM, and linear regression for simplicity and speed, but in practice a larger number of learners with greater diversity could be leveraged so that the learned models approach the true conditional means with as little data as possible. Conversely, historical data might be combined with the trial data to facilitate learning of these models. Since the AIPW estimator is unbiased regardless of the model used, including historical data in this fashion would not sacrifice type I error control. It would also be possible to leverage pre-training and transfer learning methods if deep learning were used to model the conditional means \cite{lecunn-dl, dubois-transfer}.

The divergence between the oracle and real-world AIPW estimator performance also highlights a crucial consideration for estimating population parameters for the sample size calculation. If the true population parameters were actually known, it would not behoove the analyst to use them in the power calculation because the real-world AIPW estimator would be unlikely to attain design power at the enrollment target specified with the oracle parameters. In other words, having to estimate the required population parameters from external data is actually helpful because the estimate of the average conditional variance ($\hat\kappa_0^2$) is usually closer to the amount of variance that is still ``unexplainable'' with a real-world model than the true average conditional variance ($\kappa_w^2$) is. {Indeed, our experiments with isolated GBM models were in some cases slightly under-powered, which suggests that our powering method can be sensitive to the learning curve (RMSE vs. $n$) of the chosen algorithm. Using a large, diverse ensemble is therefore recommended.} This may also have implications for the amount of historical data used to estimate $\kappa_w$ relative to the amount of data used for learning in the trial, although in our experiments we used an order of magnitude more historical than trial data and still obtained conservative sample sizes with AIPW.
Pre-training or combining historical and trial data for estimation of the conditional mean models is also an attractive avenue to address this concern in future work.

\subsection*{Data Availability}

{\small 
Certain data used in the preparation of this article were obtained from the Alzheimer’s Disease Neuroimaging Initiative (ADNI) database (\href{url}{adni.loni.usc.edu}). The ADNI was launched in 2003 as a public-private partnership, led by Principal Investigator Michael W. Weiner, MD. The primary goal of ADNI has been to test whether serial magnetic resonance imaging (MRI), positron emission tomography (PET), other biological markers, and clinical and neuropsychological assessment can be combined to measure the progression of mild cognitive impairment (MCI) and early Alzheimer’s disease (AD). For up-to-date information, see \href{url}{www.adni-info.org}.

Certain data used in the preparation of this article were obtained from the Critical Path for Alzheimer's Disease (CPAD) database. In 2008, Critical Path Institute, in collaboration with the Engelberg Center for Health Care Reform at the Brookings Institution, formed the Coalition Against Major Diseases (CAMD), which was then renamed to CPAD in 2018. The Coalition brings together patient groups, biopharmaceutical companies, and scientists from academia, the U.S. Food and Drug Administration (FDA), the European Medicines Agency (EMA), the National Institute of Neurological Disorders and Stroke (NINDS), and the National Institute on Aging (NIA). CPAD currently includes over 200 scientists, drug development and regulatory agency professionals, from member and non-member organizations. The data available in the CPAD database has been volunteered by CPAD member companies and non-member organizations.

Certain data used in the preparation of this article were obtained from the University of California, San Diego Alzheimer’s Disease Cooperative Study Legacy database.
}

\subsection*{Acknowledgments}

{\small 
The author thanks Xinkun Nie, Charles Fisher, and David Miller for helpful conversations.

Data collection and sharing for this project was funded in part by the Alzheimer's Disease Neuroimaging Initiative (ADNI) (National Institutes of Health Grant U01 AG024904) and DOD ADNI (Department of Defense award number W81XWH-12-2-0012). ADNI is funded by the National Institute on Aging, the National Institute of Biomedical Imaging and Bioengineering, and through generous contributions from the following: AbbVie, Alzheimer’s Association; Alzheimer’s Drug Discovery Foundation; Araclon Biotech; BioClinica, Inc.; Biogen; Bristol-Myers Squibb Company; CereSpir, Inc.; Cogstate; Eisai Inc.; Elan Pharmaceuticals, Inc.; Eli Lilly and Company; EuroImmun; F. Hoffmann-La Roche Ltd and its affiliated company Genentech, Inc.; Fujirebio; GE Healthcare; IXICO Ltd.; Janssen Alzheimer Immunotherapy Research \& Development, LLC.; Johnson \& Johnson Pharmaceutical Research \& Development LLC.; Lumosity; Lundbeck; Merck \& Co., Inc.; Meso Scale Diagnostics, LLC.; NeuroRx Research; Neurotrack Technologies; Novartis Pharmaceuticals Corporation; Pfizer Inc.; Piramal Imaging; Servier; Takeda Pharmaceutical Company; and Transition Therapeutics. The Canadian Institutes of Health Research is providing funds to support ADNI clinical sites in Canada. Private sector contributions are facilitated by the Foundation for the National Institutes of Health (\href{url}{www.fnih.org}). The grantee organization is the Northern California Institute for Research and Education, and the study is coordinated by the Alzheimer’s Therapeutic Research Institute at the University of Southern California. ADNI data are disseminated by the Laboratory for Neuro Imaging at the University of Southern California.

Data collection and sharing for this project was funded in part by the University of California, San Diego Alzheimer’s Disease Cooperative Study (ADCS) (National Institute on Aging Grant Number U19AG010483).
}

\bibliography{references}

\appendix

\section{Mathematical Results}

\subsection{Details of the cross-fit AIPW estimator in randomized trials}

Here we discuss the AIPW estimator and its semiparametric efficiency in the context of randomized trials. While the conclusions of this paper apply to any semiparametric efficient estimator, it may help the reader to understand semiparametric efficiency in the context of a single estimator. The details provided here are all available elsewhere in the literature or follow immediately from known results \cite{Tsiatis:2007vl, wager-stats361, rothe-rct, rosenblum-glm}. We reframe them here to provide a quick reference and starting point for further reading.

\paragraph{Preliminaries} A scalar \textit{parameter} of a distribution is a functional that ingests the distribution and returns a number. For example, consider the distribution of a scalar random variable $Y$ defined by the PDF $f_y$. The mean of $Y$ (which is a parameter) is the functional $\int y f_y(y) dy$. An \textit{estimator} of a (scalar) parameter is a function that takes data sampled from the distribution in question and returns a number which is meant to approximate the parameter. For example, an estimate of the mean of $Y$ is $\frac{1}{n}\sum_i y_i$ when $y_i$ are assumed to be draws from the distribution of $Y$. An estimator is \textit{consistent} if it recovers the true value of the parameter as the sample size grows. It is usually possible to construct many different consistent estimators of the same parameter, so we are interested in finding the one that has the smallest possible sampling variance. We call this the \textit{efficient} estimator. 

It becomes much easier to find the efficient estimator if we restrict our attention to the class of \textit{regular} and \textit{asymptotically linear} (RAL) estimators, and we actually don't lose anything by doing so. It is not important to understand the precise mathematical definition of regularity, but heuristically, a regular estimator does not have anomalously bad performance for certain special values of the parameter. Thus restricting ourselves to regular estimators is a good and sensible thing to do. Effectively all estimators which can be used in practice are regular. Moreover, among regular estimators of some parameter, a result called the Hájek-Le Cam convolution theorem guarantees that the estimator with the smallest asymptotic variance is asymptotically linear, so we lose nothing by further restricting ourselves to asymptotically linear estimators once we've excluded irregular estimators \cite{Tsiatis:2007vl}.

The definition of asymptotic linearity for an estimator $\hat\psi$ of a parameter $\psi$ is that there exists an \textit{influence function} $\phi$ such that $\sqrt{n}(\hat\psi -\psi) = \Ehat{\phi} + o_p(1)$. In other words, the estimator $\hat\psi$ behaves like an IID average of some random variable $\phi$ in large samples. Asymptotic linearity immediately implies $\sqrt{n}(\hat\psi -\psi) \rightsquigarrow N(0, \V{\phi})$ by the central limit theorem. Therefore the asymptotic variance of any asymptotically linear estimator is given by the variance of its influence function. 

Finding the most efficient RAL estimator thus boils down to finding the influence function with the smallest variance, which we call the \textit{efficient influence function} (EIF). The EIF defines a lower bound on the achievable sampling variance when estimating a parameter. It so happens that it is often possible to characterize the space of all influence functions of RAL estimators of a given parameter in a generic generative model, which makes it possible to derive the EIF. Obtaining an efficient estimator is therefore a matter of deriving the EIF for the class of RAL estimators of a parameter and then constructing an estimator that has that influence function.

\paragraph{Application}
In the semiparametric statistical model $P(Y,W,X) = P(Y|X)P(W|X)P(X)$ (with $P(Y|X)$, $P(W|X)$, and $P(X)$ free to be any distributions that satisfy mild regularity conditions), it turns out that the efficient influence function of the class of RAL estimators for the parameter $\mu_w = \E{Y|W=w}$ is $\phi_w = \frac{W_w}{\pi_w(X)}(Y-\mu_w(X)) + (\mu_w(X) - \mu_w)$. Proving this is not simple (see \citet{Tsiatis:2007vl}), but after the fact has been established all our subsequent theory requires only algebra and elementary tools from large-sample theory. 

As might be expected, the EIF for the two-dimensional parameter $\mu = [\mu_0, \mu_1]^\top$ is $\phi_\mu = [\phi_0, \phi_1]^\top$. Influence functions obey a ``chain rule'' such that if the EIF of $\psi$ is $\phi$, the EIF of $g(\psi)$ is $\nabla g^\top \psi$. For us, that means that the EIF of $\tau = r(\mu_0, \mu_1)$ is $\phi = r_0'(\mu_0, \mu_1)\phi_0 + r_1'(\mu_0,\mu_1)\phi_1$.

In a two arm randomized trial with treatment fractions $\pi_w$, one estimator of $\mu_w$ that has the efficient influence function is $\hat\mu^*_w = \Ehat{ \frac{W_w}{\pi_w}(Y-\mu_w(X)) + \mu_w(X)}$. The fact is easily verified by showing $\sqrt{n}(\hat\mu_w -\mu_w) = \Ehat{\phi_w} + o_p(1)$ with an application of the central limit theorem. An application of the delta method shows that $\hat\tau^* = r(\hat\mu_0^*, \hat\mu_1^*)$ attains the asymptotic variance $\nu^2 = \V{\phi}$ and is therefore efficient. We'll call this the ``oracle estimator''.

Unfortunately, the oracle estimator $\hat\tau^*$ is infeasible in practice because the conditional means $\mu_w(X)$ are not known. However, it turns out that estimates can be substituted without sacrificing the optimality properties. Let
$
\hat\mu_w^{(k)} 
= 
\Ehat{ \frac{W_w}{\pi_w}(Y-\hat\mu_w^{(-k)}(X)) + \hat\mu_w^{(-k)}(X) \ \big| \  k(i) = k}
$
. This is the marginal mean estimated from the $k$th fold of data using the conditional mean function \textit{estimated} from the rest of the data. Let $\hat\mu_w^{*(k)}$ be the oracle equivalent that uses the \textit{true} conditional mean $\mu_w(X)$. Clearly, $\hat\mu_w = \sum \frac{n^{(k)}}{n} \hat\mu_w^{(k)}$ where $n^{(k)}$ is the number of observations in fold $k$ (and similarly for $\hat\mu_w$). If we can show that $\sqrt{n}(\hat\mu_w^{(k)} - \hat\mu_w^{*(k)}) \rightarrowp 0$, then Slutsky's theorem and the delta method imply that $\hat\tau$ has the same asymptotic properties as $\hat\tau^*$, i.e. $\sqrt{n}(\hat\tau - \tau) \rightsquigarrow N(0, \nu^2)$. In other words, since the oracle estimator is efficient with a known asymptotic variance, the feasible estimator is also efficient and has the same asymptotic variance. It turns out that if we assume that 
$
\E{
\left(
    \hat\mu_w^{(-k)}(X) - \mu_w(X)
\right)^2
} 
\rightarrow 0
$
(a very weak condition), then it is possible to show $\sqrt{n}(\hat\mu_w^{(k)} - \hat\mu_w^{*(k)}) \rightarrowp 0$ as desired. 

We begin by deriving an expression for the difference between the oracle and feasible estimators:

\begin{equation}
\begin{split}
\hat\mu_w^{(k)} - \hat\mu_w^{*(k)}
&=
\Ehat{
\left(
     \frac{W_w}{\pi_w}(Y-\hat\mu_w^{(-k)}(X)) + \hat\mu_w^{(-k)}(X)
\right) 
    -
\left(
    \frac{W_w}{\pi_w}(Y-\mu_w(X)) + \mu_w(X)
\right)
\ \bigg | \  k(i) = k
}
\\
&= 
\Ehat{
\left( 1 - \frac{W_w}{\pi_w} \right)
\left( \hat\mu_w^{(-k)}(X) - \mu_w(X) \right)
\ \bigg | \  k(i) = k
}
\\
&=
\frac{1}{n^{(k)}}
\sum_{i \in \mathcal I_k}^{n^{(k)}}
\left[
\left( 1 - \frac{W_{w,i}}{\pi_w} \right)
\left( \hat\mu_w^{(-k)}(X_i) - \mu_w(X_i) \right)
\right]
\end{split}
\label{eq:oracle-diff}
\end{equation}

where in the last line all we've done is expand the $\Ehat{\cdot}$ notation and introduce $\mathcal I_k = \{i:k(i)=k\}$. Notice that if we condition on the dataset $\mathcal I_{(-k)}$, the estimated function $\hat\mu_w^{(-k)}(\cdot)$ becomes fixed because it does not depend on the data in fold $k$. Therefore 
$
\E{
\hat\mu_w^{(k)} - \hat\mu_w^{*(k)} 
\ \big| \ \mathcal I_{(-k)}
} = 0
$ 
because of the conditional independence between $\left( 1 - \frac{W_{w,i}}{\pi_w} \right)$ and $\left( \hat\mu_w^{(-k)}(X_i) - \mu_w(X_i) \right)$ in fold $k$ and the fact that $\E{\left( 1 - \frac{W_{w,i}}{\pi_w} \right)}=0$. This, in turn, implies that 
$
\E{
\left(\hat\mu_w^{(k)} - \hat\mu_w^{*(k)} \right)^2
\ \big| \ \mathcal I_{(-k)}
}
=
\V{
\hat\mu_w^{(k)} - \hat\mu_w^{*(k)} 
\ \big| \ \mathcal I_{(-k)}
}
$,
which we will use shortly. Moreover, the terms within the sum of eq. \ref{eq:oracle-diff} are all IID conditional on $\mathcal I_{(-k)}$ so we can pass the variance through the sum (and gain a $1/n^{(k)}$ in the process):

\begin{equation}
\begin{split}
\V{
\hat\mu_w^{(k)} - \hat\mu_w^{*(k)} 
\ \big| \ \mathcal I_{(-k)}
}
&=
\frac{1}{(n^{(k)})^2}
\sum_{i \in \mathcal I_k}^{n^{(k)}}
\V{
\left( 1 - \frac{W_{w,i}}{\pi_w} \right)
\left( \hat\mu_w^{(-k)}(X_i) - \mu_w(X_i) \right)
\ \big| \ \mathcal I_{(-k)}
}
\\
&=
\frac{1}{n^{(k)}}
\left( \frac{1-\pi_w}{\pi_w} \right)
\E{
\left( \hat\mu_w^{(-k)}(X_i) - \mu_w(X_i) \right)^2
\ \big| \ \mathcal I_{(-k)}
}
\\
\end{split}
\end{equation}

Our plan is to show
$
\sqrt{n}(\hat\mu_w^{(k)} - \hat\mu_w^{*(k)}) \overset{L^2}{\rightarrow} 0
$
, which implies
$
\sqrt{n}(\hat\mu_w^{(k)} - \hat\mu_w^{*(k)}) \rightarrowp 0
$
because $L^2$ convergence implies convergence in probability. By the definition of $L^2$ convergence, that means we must show $\E{n\left(\hat\mu_w^{(k)} - \hat\mu_w^{*(k)}\right)^2} \rightarrow 0$, which we can do by using what we've derived above:

\begin{equation}
\begin{split}
\E{n\left(\hat\mu_w^{(k)} - \hat\mu_w^{*(k)}\right)^2}
&=
n\E{\E{
\left(\hat\mu_w^{(k)} - \hat\mu_w^{*(k)}\right)^2 
\ \big | \ \mathcal I_{(-k)}
}}
\\
&=
n\E{\V{
\hat\mu_w^{(k)} - \hat\mu_w^{*(k)}
\ \big | \ \mathcal I_{(-k)}
}}
\\
&=
\frac{n}{n^{(k)}}
\left( \frac{1-\pi_w}{\pi_w} \right)
\E{
\E{
\left( \hat\mu_w^{(-k)}(X_i) - \mu_w(X_i) \right)^2
\ \big| \ \mathcal I_{(-k)}
}
}
\\
&=
\frac{n}{n^{(k)}}
\left( \frac{1-\pi_w}{\pi_w} \right)
\E{
\left( \hat\mu_w^{(-k)}(X_i) - \mu_w(X_i) \right)^2
}
\\
\end{split}
\end{equation}

This must converge to $0$ because $n$ and $n^{(k)}$ grow in proportion to each other and the expectation 
$
\E{
\left( \hat\mu_w^{(-k)}(X_i) - \mu_w(X_i) \right)^2
}
$
converges to 0 by our mean-square consistency assumption (eq. \ref{eq:mse-norm}).

By the arguments above, this is enough to establish the asymptotic normality of our estimator with the efficient asymptotic variance: $\sqrt{n}(\hat\tau - \tau) \rightsquigarrow N(0, \nu^2)$.

Similar arguments are required to show that the plug-in variance estimator is consistent (when cross-fitting is used). Some care is required to address the terms $r_w'(\hat\mu_0, \hat\mu_1)$, etc., but the elementary tools of asymptotic statistics suffice.

\subsection{Proof of theorem \ref{thm:asymptotic-variance}}

Beginning with eq. \ref{eq:asymptotic-variance}, we have that 

\begin{equation}
\begin{split}
    \nu_*^2 &= \V{
        r'_0 \phi_0 +
        r'_1\phi_1
    } \\
    & = 
        r_0'^2\V{\phi_0} + 
        r_1'^2\V{\phi_1} + 
        2 r_0'r_1'\C{\phi_0, \phi_1}
\end{split}
\label{eq:asymptotic-variance-expansion}
\end{equation} 

from expansion of the variance. We will use several tricks to analyze these terms. The first is that $W_wY = W_wY_w$, by the fact that $Y = Y_0W_0 + Y_1W_1$ (note $W_w^2 = W_w$ and $W_0W_1 = 0$, which we will also use). Secondly, $W_w \perp Y_w, \hat \mu_w(X)$ by unconfoundedness, allowing for factorization of expectations. 

These tricks and a few lines of algebra show that the variances in the first and second terms above (eq. \ref{eq:asymptotic-variance-expansion}) are 

\begin{equation}
\begin{split}
    \V{\phi_w} 
    &= \frac{1-\pi_w}{\pi_w}\E{\sigma^2_w(X)}
    + \V{Y_w} 
    \\
    &= \frac{1-\pi_w}{\pi_w} \kappa^2_w
    + \sigma_w^2
    \\    
\end{split}
\end{equation}

where $\sigma_w^2(X) \equiv \V{Y_w|X}$ is the conditional variance function in treatment arm $w$. In the last line we define $\kappa^2_w. \equiv \E{\sigma^2_w(X)}$ (the average conditional variance) and $\sigma^2_w \equiv \V{Y_w}$ (the marginal variance) to simplify notation. 

Similar manipulation reduces the covariance term in eq. \ref{eq:asymptotic-variance-expansion} to

\begin{equation}
\begin{split}
\C{\phi_0, \phi_1} 
&=
\C{\mu_0(X), \mu_1(X)}
\\  &=
\text{Corr}[\mu_0(X), \mu_1(X)] \sqrt{\V{\mu_0(X)} \V{\mu_1(X)}}
\\ &=
\gamma \sqrt{(\sigma_0^2 - \kappa^2_0)(\sigma_1^2 - \kappa^2_1)}
\end{split}
\end{equation}

by exploitation of the law of total variance 
$
\underbrace{\V{Y_w}}_{\sigma_w^2} = \underbrace{\E{\sigma^2_w(X)}}_{\kappa^2_w} + \V{\mu_w(X)}
$ 
and definition of 
$
\gamma \equiv \text{Corr}[\mu_0(X), \mu_1(X)]
$.
Assembling the above and making explicit the known signs of $r'_w$ gives

\begin{equation}
\begin{split}
    \nu_*^2 
    & = 
        r_0'^2 \left(
            \frac{\pi_1}{\pi_0} \kappa^2_0 + \sigma_0^2
        \right) +
        r_1'^2 \left(
            \frac{\pi_0}{\pi_1} \kappa^2_1 + \sigma_1^2
        \right) - 
        2 |r_0'r_1'| \gamma \sqrt{(\sigma_0^2 - \kappa^2_0)(\sigma_1^2 - \kappa^2_1)}
\end{split}
\end{equation} 

\section{Additional Simulation Results}

\begin{figure}[h]
    \centering
    \includegraphics[width=1\textwidth]{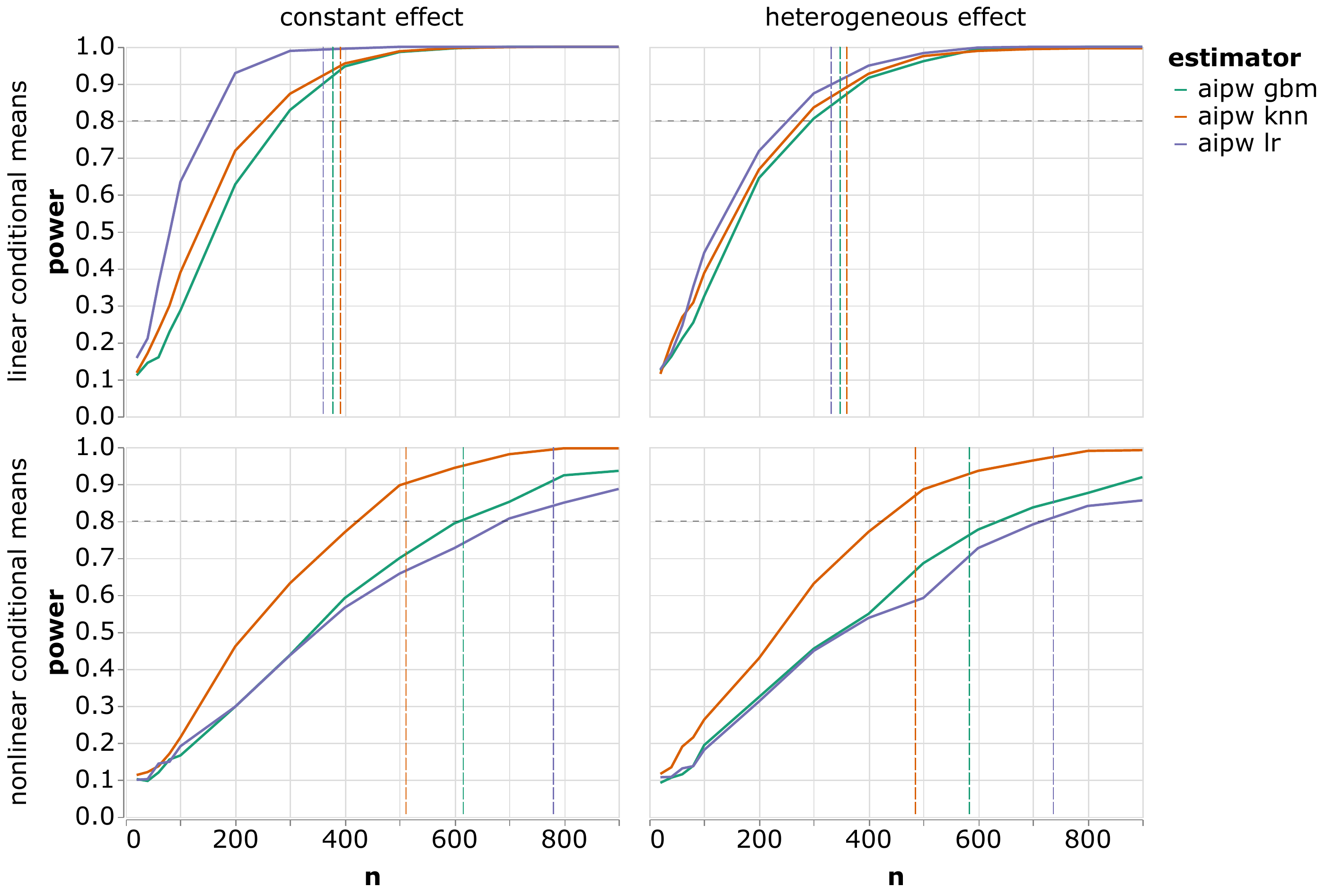}
    \caption{
    {
    Empirical power and prospectively-calculated enrollment targets for AIPW estimators with different learners used to estimate the conditional means. Visual elements are as in figures \ref{fig:power} and \ref{fig:power-M}.
    }
    }
    \label{fig:power-M-app}
\end{figure}

\end{document}